\documentclass[preprint,proceedings,nonatbib]{rmaa}
\usepackage{natbib}
\bibpunct{(}{)}{;}{a}{}{,}

\usepackage{graphicx}
\usepackage{booktabs}
\usepackage{aas_macros}
\usepackage{paralist}

\SetYear{2003}
\SetConfTitle{Compact Binaries in the Galaxy and beyond}

\title{AM~CV\lowercase{n} systems as optical, X-ray and GWR sources}  

\author{
  L. Yungelson,\altaffilmark{1} 
  G. Nelemans,\altaffilmark{2}
  and S. F. Portegies Zwart\altaffilmark{3}}

\altaffiltext{1}{Institute of Astronomy, RAS, Moscow, Russia.}
\altaffiltext{2}{Institute of Astronomy, U. Cambridge, UK.}
\altaffiltext{3}{U. Amsterdam, the Netherlands.}

\shortauthor{Yungelson, Nelemans, \& Portegies Zwart}
\shorttitle{AM~CV\lowercase{n}~systems}

\fulladdresses{
\item L. Yungelson: Institute of Astronomy of the Russian Academy of Sciences, 
 48 Pyatnitskaya Str., 109017 Moscow, Russia 
  (\email{lry@inasan.rssi.ru}).
\item G. Nelemans: Institute of Astronomy, University of Cambridge, Madingley Road, Cambridge CB3 0HA, UK (\email{nelemans@ast.cam.ac.uk}).
\item S.F.  Portegies Zwart: Astronomical Institute ``Anton Pannekoek'' and Section
  Computational Science, University of Amsterdam,
  Kruislaan 403, NL-1098 SJ Amsterdam, the Netherlands  (\email{spz@science.uva.nl}).
}

\listofauthors{L. Yungelson, G. Nelemans, \& S. F. Portegies Zwart}
\indexauthor{Yungelson, L.}
\indexauthor{Nelemans, G.}
\indexauthor{Portegies Zwart, S. F.}

\abstract{  We discuss the model for the Galactic sample of the
AM~CVn systems with $P_{\mathrm{orb}} \leq 1500$\, s that can be detected in the optical
and/or X-ray bands and may be resolved by the gravitational waves detector
\textit{LISA}. 
At $3 \lesssim P \lesssim 10$\,min
all detectable systems are X-ray sources. At $P \gtrsim 10$\,min most systems
are  only detectable in the $V$-band. About 30\% of the X-ray sources is
also detectable in the $V$-band.
About 10,000 AM~CVn
systems might be resolved by \textit{LISA}; this is  comparable to the number
of detached double white dwarfs that can be resolved. Several hundreds of AM~CVn \textit{LISA} sources might be also detectable in the
$V$- and/or X-ray bands.
 }

\resumen{ }

\addkeyword{BINARIES: CLOSE}
\addkeyword{GRAVITATIONAL WAVES}
\addkeyword{WHITE DWARFS}

\begin{document}
% Typeset article header
\maketitle

\section{INTRODUCTION}
\label{sec:intro}

AM~CVn stars are short-period binaries ($P \lesssim 1$\,hr) that are observed at optical wavelengths as faint blue and variable objects. Their optical variability is of dwarf Novae or nova-like type. Some  AM~CVn stars are also observed as X-ray sources. Currently, the class includes 13 confirmed and candidate objects. Note, that two
candidates have the shortest orbital periods and were first discovered
as X-ray sources. 
Observations of AM~CVn systems were recently reviewed,  e.g., by \citet{sol03}.
A discussion of the model of
the population of AM~CVn stars can be found in \citet{nyp03}.

AM~CVn stars are thought to be 
 semidetached binaries
with both
components being white dwarfs; evolution of the system is driven by the loss of orbital angular momentum via radiation of gravitational waves \citep{pac67}.
The total Galactic number  of AM~CVn stars
is estimated as $\sim 3.7 \cdot 10^7$ \citep{nyp03}, but the state of understanding of their formation  and uncertainties in, e.g., the star formation history, the initial mass function of stars and the parameters entering population synthesis  make this estimate uncertain by 
a factor $\sim 10$ \citep{nyp01a,nyp03}.

We consider the model for the subset of shortest period AM~CVn stars ($P \leq 1500$\, s) that
could be detected by the gravitational wave
(GWR) detector \textit{LISA} and focus on the systems that may have optical and/or X-ray
counterparts. At $P \gtrsim 1500$\,s AM~CVn stars do not emit interesting amounts of energy in the GW band and optical
radiation from the disc completely dominates emission of the star.

The model of the population of AM~CVn stars is obtained by population synthesis code \textsf{SeBa} 
\citep{pv96,nyp03}. The results we
discuss represent one possible random realization of the model, so all numbers given
are subject to Poisson noise.

\section{OPTICAL, X-RAY and GW EMISSION OF AM~CV\lowercase{n} STARS}
\label{sec:em}

Some of the AM~CVn stars, are so compact that they are in  the ``direct impact'' phase when 
the accretion disk is absent. Thus,
there are four emission sites: the accretion disc,
the impact spot in the
direct impact accretors, the donor and the accretor.

If the disc is present, we assume that half of the energy released by accretion is radiated by the disc in the optical band, while the rest is emitted as soft X-rays from the boundary layer. For the estimate of optical emission
we apply a single-temperature disc model \citep{wade84}.  For the X-ray emission of boundary layer we use the model of \citet{pringle77}.  

For direct impact systems we assume that whole accretion luminosity is radiated as a blackbody within the fraction $f=0.001$ of the total surface of the dwarf \citep{mar_ste02}.

Emission of the donor and accretor comes from the cooling which we approximate using models of \citet{hansen99}. We neglect irradiation effects and tidal or compressional heating.     

Interstellar reddening is taken into account using the \citet{san72} model. For the estimate of absorption in  X-rays, hydrogen column density is modeled after  \citet{pre_sch95}. 

We assume that AM~CVn systems emit GW as point sources. 
The strain amplitude  is evaluated
by standard formalism \citep[e.g.,][]{eis87}.

\begin{figure}[t!]
\resizebox{\columnwidth}{!}{\includegraphics[]{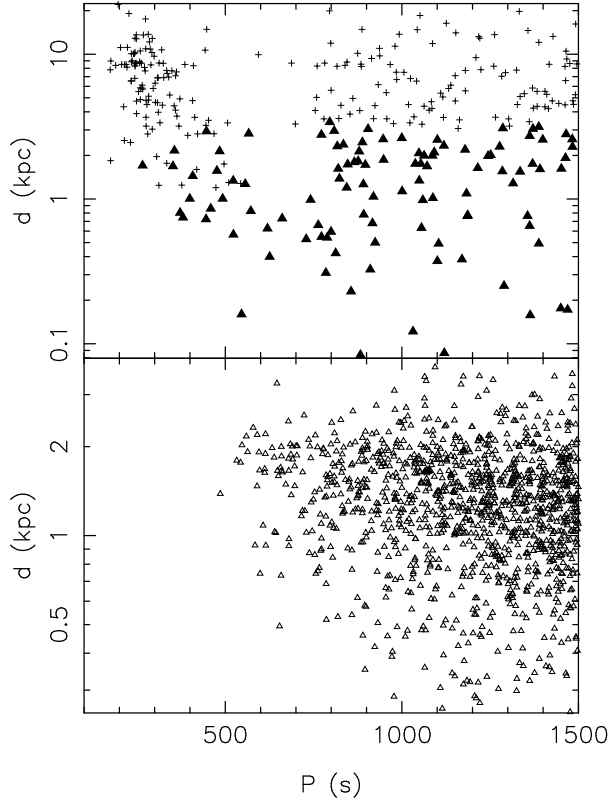}}
\caption[]{Top panel: Period vs. distance distribution of systems detectable only in X-rays (pluses), and systems that are detectable in the optical and
  X-ray band (filled triangles). Bottom
    panel: Systems only detectable in the optical band.  }
\label{fig:Pd}
\end{figure}

\section{RESULTS }
\label{sec:res}

In Figure~\ref{fig:Pd} we show the model sample of the systems that would be
detectable in the optical and/or X-ray bands.
We limit the ``optical'' sample by  
$V_{\it lim} = 20$\,mag, which is close to the
$V_{\it lim}$ of 
observed AM~CVn stars. The  ``X-ray'' sample is comprised by the systems that have a flux in the 0.1 -- 2.4 keV (\textit{ROSAT}) band higher than 
$10^{-13}$~erg~s$^{-1}$~cm$^{-2}$.

The top panel of Fig.~\ref{fig:Pd} shows 220 systems detectable in X-rays only (pluses) and the additional 106 systems that 
are also detectable in the $V$-band (filled triangles).
 There are 75 systems with optical emission from a disc and 28 systems close to the Earth,  in which the 
donor stars can be seen as well as the discs.
About half of X-ray systems are in the direct impact phase.  Note, at the shortest periods a significant fraction of X-ray systems 
are detectable even close to the Galactic center. 

The bottom panel of Fig.~\ref{fig:Pd} shows 1230 ``conventional'' AM~CVn
systems, detectable only by the optical emission, mainly from their
accretion disc. Of this population 169 ones
closest to the Earth also have a visible donor. 
The ``optical'' systems with periods between 1000 and 1500 s are expected to have unstable discs \citep{to97}; this may enhance the probability of their discovery.

Figure~\ref{fig:Pd} clearly predicts the existence of a significant population of  systems  that might
be detectable only as X-ray sources. Some have periods as short as
three minutes. About one third of the X-ray sources is also detectable
in the optical band. This prediction is sustained by discovery
of sources that are  detectable in X-rays and (just)
in the optical band.

\begin{figure*}[t!]
\resizebox{2\columnwidth}{!}{\includegraphics[angle=-90,clip]{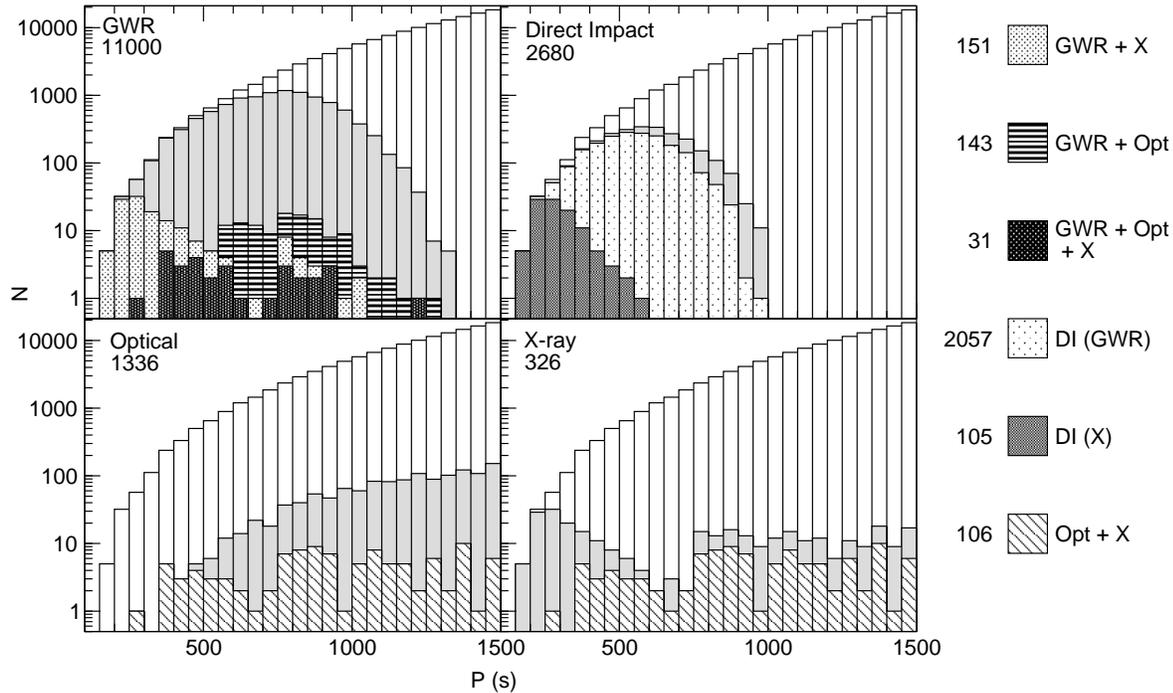}}
\caption[]{%Histograms of the population of
 Short-period AM~CVn
  systems, subdivided in different types. Each panel shows
  the \emph{total} population as the white histogram. The
  top left panel shows the systems that can be resolved by
  \textit{LISA} in grey and they are subdivided in the ones that have
  optical counterparts (GWR + Opt), X-ray counterparts (GWR + X) and
  both (GWR + Opt + X).  The top right panel shows the
  systems that are in the direct impact phase of accretion in grey
  and it is subdivided in GWR and X-ray sources. The
  bottom two panels show (again in grey) the populations that are
  detectable in the optical band (left panel) and the X-ray band
  (right panel). The distribution of sources detectable both in
  optical and X-ray bands is shown as hatched bins in both lower
  panels (Opt + X). 
 }
\label{fig:hist_all}
\end{figure*}
\nopagebreak
Previous studies of GW emission of the AM~CVn systems \citep[e.g.][]{hb00,
nyp01} have found that they hardly
contribute to the GW background noise,
even though at $\nu = 0.3 - 1$\,MHz they outnumber the detached double white
dwarfs. This is because at these $\nu$
their chirp mass $\mathcal{M}$ is
well below that of a typical detached system. But it was overlooked that at higher $\nu$, where the number of AM~CVn systems is much smaller,
their $\mathcal{M}$ is similar to that of the detached systems from which
they
descend. Our model shows that, out of the total population of $\sim 140,000$ AM~CVn stars with $P\leq1500$\,s, for $T_{\mathrm{obs}}=1$\,yr, \textit{LISA} may be expected  
to resolve  $\sim$11,000 double white
  dwarfs at $S/N \geq 5$ as well as $\sim$ 11,000 AM~CVn systems at $S/N \geq 1$ (or $\sim 3000$ at  $S/N \geq 5$). 

In Figure~\ref{fig:hist_all} we show the distributions vs.
orbital periods 
for the total number of AM~CVn systems with $P \leq 1500$ s and for AM~CVn-\textit{LISA} sources that have optical and/or X-ray counterparts,
direct impact systems that are GWR and X-ray sources, and the total
number of systems that are optical and/or X-ray sources. The interrelations between numbers of sources emitting in different wave bands are shown
in the legend at the right of the Figure. 

It is clear from Fig.~\ref{fig:hist_all} that \textit{LISA} is
expected to be very effective in detecting the short-period AM~CVn
systems.
About 20\% of them will be in the direct impact phase and at short periods might be detectable in X-rays.
Towards longer periods the systems detectable in
the optical band take over. 

The GW signal from a monochromatic binary is characterized by seven independent parameters. Therefore in the data analysis all parameters, except the amplitude $h$, must
be estimated using a fitting technique for which an initial guess is needed
\citep{hell03}. Information from observations in the optical and X-ray bands
will facilitate these  guesses.
On the other hand, e.g., inclinations of orbits and chirp masses obtained  by GW observations would provide information for the estimates of parameters of AM~CVn stars.

L.Y. acknowledges financial support from CONACYT 34521-E and RFBR 03-06-16254 grants.

\end{document}